\documentclass[fleqn]{article}
\usepackage{espcrc2}

\usepackage{graphicx}

\newcommand{\AmS}{{\protect\the\textfont2
  A\kern-.1667em\lower.5ex\hbox{M}\kern-.125emS}}

\title{Population synthesis modeling of the X-ray background with 
genetic algorithm -- based optimization method}

\author{Alexander V. Halevin\address[ONU]{Department of Astronomy, Odessa National 
University, \\
T.G.Shevchenko park, 65014, Odessa, Ukraine,\\ halevin@astronomy.org.ua}%
}
       
\begin{document}

\begin{abstract}
We present population synthesis modeling of the X-ray background 
with genetic algorithm -- based optimization method. In our models 
the best fit could be achieved for lower values of high-energy 
exponential cut-off ($\approx 170~keV$) and larger amount of the 
highly obscured ($\log N_{H}=25.5$) AGNs. \vspace{1pc} 
\end{abstract}

\maketitle

\section{Introduction}

The cosmic X-ray background (XRB) above $\sim1$ keV is known to be
produced by integrated emission of discrete sources 
\cite{Mush,Has02,Mo}. The resent best reviews in this area are 
\cite{Gil03} and \cite{Has03}. XRB synthesis models are usually 
based on the so-called unification scheme for AGN \cite{Ant}, 
which explains the different observational appearances by the 
orientation of accretion disk and molecular torus surrounding the 
nucleus. The intersection of the line of sight with the torus 
determines a type 2 AGNs, and the direct observation of the 
nucleus identifies a type 1 AGNs. 

The last works in the area of population synthesis modeling used
assumption about extra quantity of absorbed AGNs at high redshifts
\cite{Wil,Gil99a,Pom,Gil01,Mor}. These models give good approach 
to the exist observations in the energy range of $1\div100$ kev. 

However, there are many unresolved yet problems exist, like as a
role of the soft X-ray excess in AGN spectra at high redshifts,
behaviour of the AGN luminosity function \cite{Has03}, possible 
flattening of the spectra of AGNs at high redshifts and a role of 
the high luminosity obscured AGNs. Furthermore, we do not have yet 
detailed information about an exponential cut-off at high 
energies. 

Here we present population synthesis models of the X-ray 
background in order to investigate some additional conditions to 
obtain the best fit. 

\section{Fitting method}

In our work we have tested one of the newest effective fitting 
techniques: a genetic algorithm -- based optimization \cite{Char}. 
This method implements the Darvin's natural selection law for the 
mathematical problems. 

The main steps of this technique are:

1. Constructing a random initial population of the sets of the
model parameters and evaluating the fitness of its members. 2. 
Constructing a new population by breeding selected individuals 
from the old population. 3. Evaluating the fitness of each member 
of the new population. 4. Replacing the old population by the new 
population. 5. Test convergence: unless fittest phenotype matches 
target phenotype within tolerance, goto 2. 

The breeding consists of the several steps such as crossover (as a
result the offspring obtains the properties of the both its
parents) and mutation (which allows to probe the alternative sets
of the model parameters. This option especially important, when
population members become practically identical).

This method already was implemented as one of the XSpec fitting 
methods and now often used for different complicated fitting 
problems.

\section{Description of the model}

As for any population synthesis model, the resulting spectrum was
calculated as a mix of the AGN spectra, which are typical for the
different classes of AGNs, integrated over redshift and
luminosity. To avoid a contribution of the Galactic diffuse
radiation component, we have modeled the energy range above 1.5 
keV only. 

Thus, following \cite{Com}, for AGN spectra we have the next 
expressions: 

\begin{equation}
F(Quasars~1)\propto E^{-\alpha_{h}}*\exp(-\frac{E}{E_{c}})
\end{equation}

\begin{equation}
F(Seyferts~1)\propto E^{-\alpha_{h}}*\exp(-\frac{E}{E_{c}}) +
F_{refl}(E)
\end{equation}

\begin{equation}
F(Type~2)\propto F(Type~1)*\exp(-\sigma_{E}*N_{H})
\end{equation}

\noindent where the hard energy index $\alpha_{h}=0.9$. For the
exponential cut-off as a base value we have used $E_{c}=320~keV$,
but in different models we have tried another values.

The term $F_{refl}(E)$ represents the Compton reflection component
by the accretion disk and has been evaluated following \cite{Mag} 
assuming inclination angle of $60^{\circ}$. 

For obscured (type 2) AGNs, photoelectric absorption cross
sections for given hydrogen column $N_{H}$ were calculated
following \cite{Morr}. As the base distribution of equivalent 
hydrogen column densities $N_{H}$ was taken from \cite{Ris} (RMS 
hereafter). Following \cite{Mai}, the local ratio of absorbed and 
unabsorbed AGNs is $R=4.0\pm0.9$. To evaluate this ratio with 
redshift, we have used the next formula 

\begin{eqnarray*}
R(z)=\left(R(0) + \Delta 
R(\infty)*\left(1-\frac{1}{1+z/z_{r}}\right)\right)\times\\ \quad 
\times\exp(-z/z_{e}) 
\end{eqnarray*}

\noindent which in contrast to the "power law \& constant" form of 
\cite{Gil01} has continuous first derivative and is convenient to 
represent the ratio dependence from the redshift. Here 
$(R(0)+\Delta R(\infty))$ is the ratio at "infinity" and $z_{r}$ 
is a distance for R(z) to get value $R(0)+\frac{\Delta 
R(\infty)}{2}$. $z_{e}$ is an exponential cut-off scale, the same 
as in \cite{Pom}. 

During integration for the X-ray luminosity function we have used
expression in the form presented by \cite{Miy00}. For all our 
models the Hubble constant given by $H_{0}=50~km~s^{-1}~Mpc^{-1}$. 

We also computed contribution of the clusters of galaxies to the
overall background spectrum, adopting the temperature distribution
taken from \cite{Dav} and cluster luminosity function of
\cite{Ebel}.

Galactic hydrogen column was taken as $N_{H}^{gal}=10^{20}~
cm^{-2}$.

To fit our models we have used kindly provided data from different 
missions like as HEAO-1 A4 LED and A2 HED detectors 
\cite{Grub92,Grub99,Bol}, ROSAT PSPC and ASCA GIS 
\cite{Miy98,Miy03}. 

The main parameters which describe our implementation of the
genetic algorithm are population number (100 members for the model 
with 3 free parameters and 200 members for the case of 4 free 
parameters), variable mutation rate (initial value is 0.005) and 
high selection pressure regime for a breeding of the population 
members. Furthermore, we have used so-called "elitism", when the 
fittest generation member is copied without alteration into the 
next generation. A good algorithm convergence we achieve after 
40--50 iterations. 

\section{Results and discussion}

In the present work we realized several kinds of models. A simple
model $A$ was calculated using equation 4 with $\Delta R(\infty)$, 
$z_{r}$ and $z_{e}$ as free parameters. The best fit results 
presented in the Table 1. The corresponded XRB fit and $R(z)$ 
curve one can see in Fig.1 and Fig.2. 

For the model A we have the same rapid increase of the $R(z)$ with 
redshift, as in work of \cite{Gil01} and significantly much more 
slower exponential decay than found by \cite{Pom}. We consider 
that such rapid increase of $R(z)$ can not be real because it 
makes an observer as a particular person. The most probable 
mistake could be using 'as is' the distribution of equivalent 
hydrogen column densities of RMS. In spite of very detailed 
calculations of the selection effects, their sample based on 
optical identifications using emission lines and, hence, can lose 
some highly obscured AGNs. In reality the distribution could 
shifts towards high values of hydrogen column densities. 

In order to investigate this possibility in our model B we have 
used as an additional parameter the quantity of AGNs with 
$\log{N_{H}}=25.5$. In the Table 1 the last parameter $n_{25.5}$ 
was calculated as a fraction of all population of type 2 AGNs in 
units of the RMS fractional population value (they have used 
$n_{25.5}\approx0.26$). Then new local ratio of absorbed and 
unabsorbed AGNs is calculated as 
$4.0\times(1-n_{25.5}^{old}+n_{25.5}^{new})$. You can see that our 
new value of $n_{25.5}$ is distinctly higher than used before. For 
new model values the XRB spectrum is better fitted below 50 keV 
but is worse above 50 keV. At the same time the $R(z)$ function 
becomes more smooth. The distribution of hydrogen column densities 
for the different cases one can see in the Fig. 3. 

The main problem of our fitting curves is high bias to the data 
above 50 keV. In reality if we analyse the data up to 400 keV it 
is obvious that we do not have clear exponential cut-off profile. 
Assuming that data below 100 keV represents the 'true' exponential 
cut-off we have tried to make the model C with $E_{C}=160~keV$. 
One can see that in this case we achieve practically ideal 
approach to the observations, although it demands to increase the 
$n_{25.5}$ parameter up to 0.93.

\begin{center} 
Table 1. The best fit parameters for different 
models.\nopagebreak\\ 
\begin{tabular}{|c|cccc|}\hline
model&$\Delta R(\infty)$&$z_{r}$&$z_{e}$&$n_{25.5}$ 
\\\hline
  A & 3.34 & 0.11 & 2.49 & - \\
  B & 1.21 & 3.01 & 4.46 &  0.47 \\
  C & 2.64 & 0.47 &  2.96 & 0.93 \\ \hline
\end{tabular}\\
\end{center}

In spite of approximate character of our models we have to 
conclude the presence of a large amount of undetected highly 
obscured AGNs (with $\log N_{H}=25.5$). I our future models we are 
going to use more precise simulations of the spectra of absorbed 
AGNs and the latest results obtained for the properties of the 
luminosity function.\\


{\it Acknowledgements. \rm Author is thankful to Duane Gruber, 
Elihu Boldt, Takamitsu Miyaji and Keith Gendreau for providing of 
the spectral data from different missions and/or useful comments. 
Special thanks to G$\rm \ddot u$nther Hasinger and Roberto Gilli 
for their exciting article "The Cosmic Reality Check".\\}

\begin{figure*}[t]
\vspace{0cm} \hspace{0cm} 
\resizebox{\hsize}{!}{\includegraphics{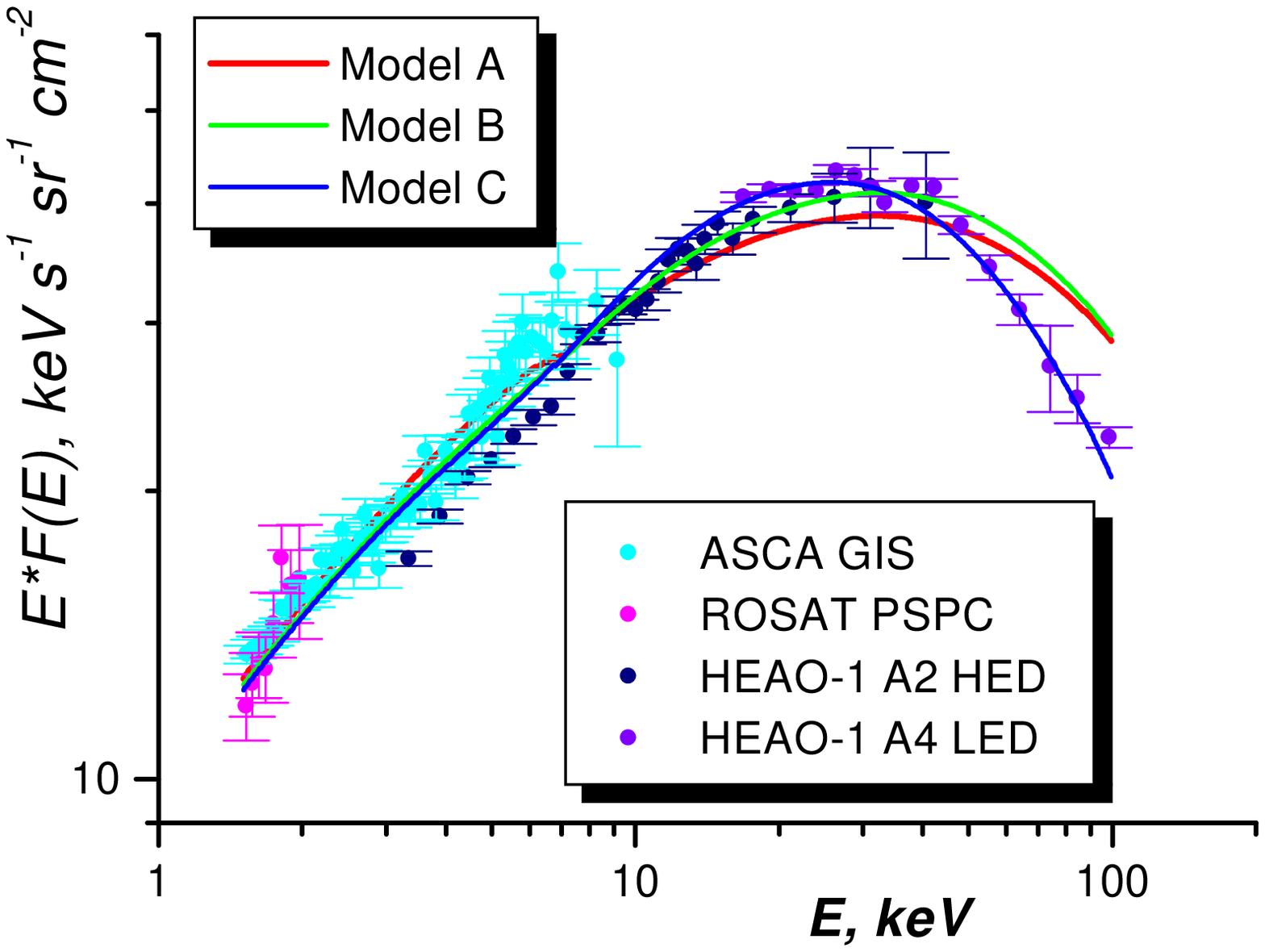}} \caption{XRB 
data from different missions and our model fits.} \label{fig1} 
\vspace{0cm} 
\end{figure*}

\begin{figure*}[b]
\vspace{0cm} \hspace{0cm} 
\resizebox{\hsize}{!}{\includegraphics{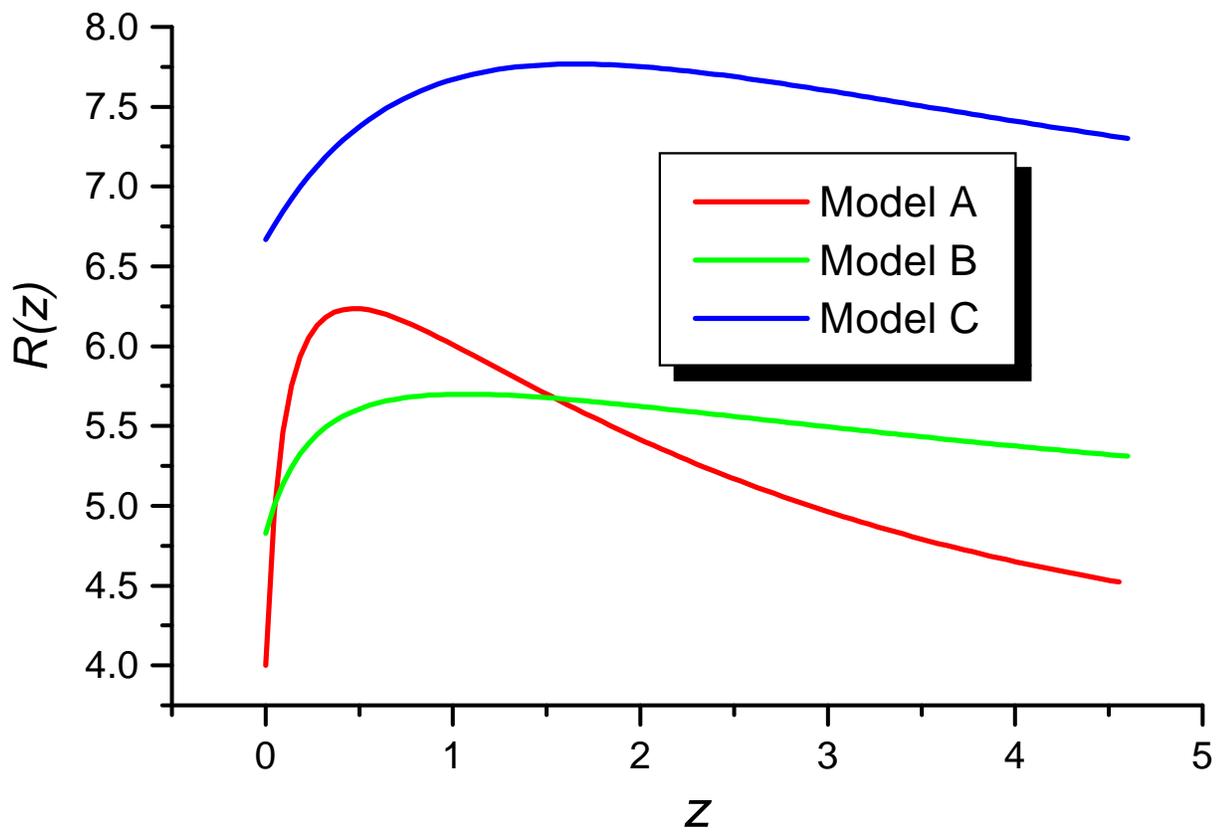}} \caption{The 
ratio of absorbed and unabsorbed AGNs for different redshifts.} 
\label{fig1} \vspace{0cm} 
\end{figure*}

\begin{figure*}[b]
\vspace{0cm} \hspace{0cm} 
\resizebox{\hsize}{!}{\includegraphics{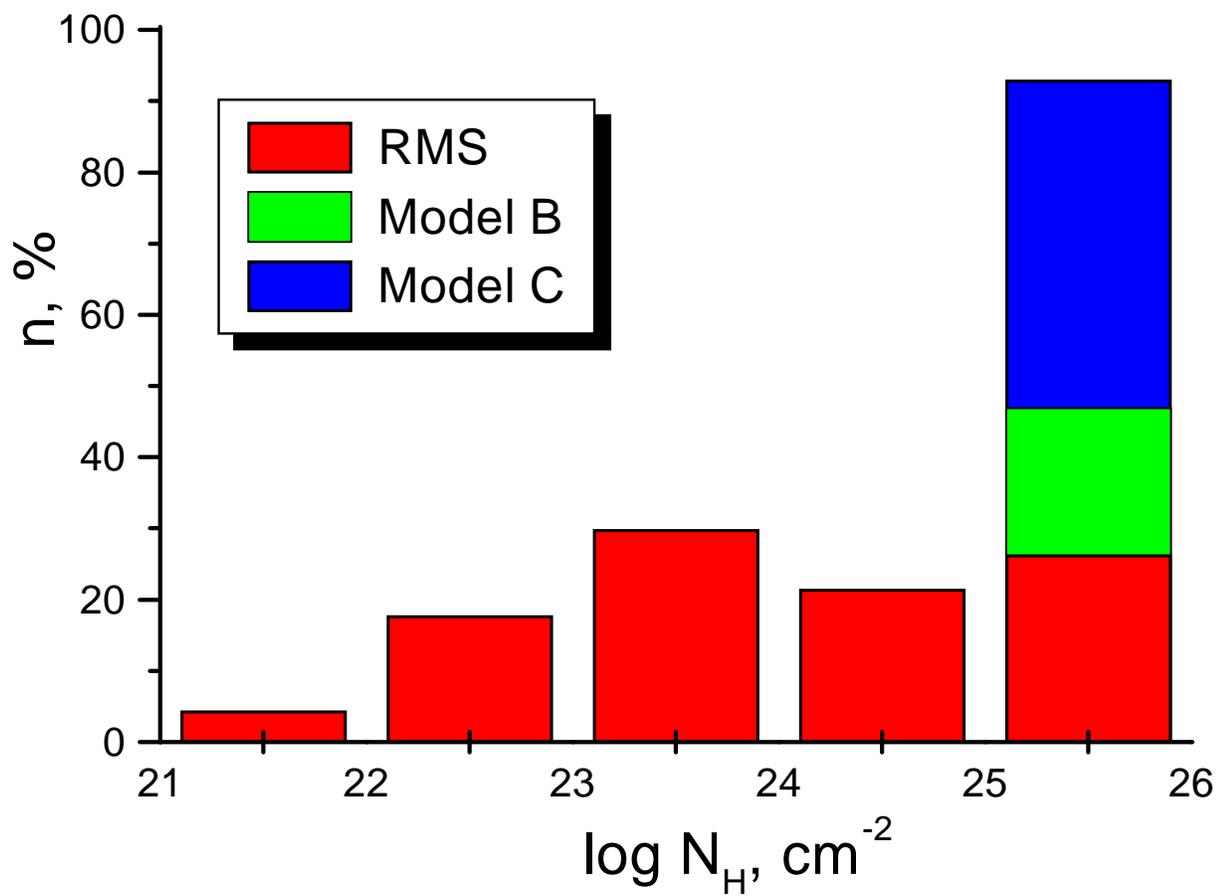}} \caption{ The 
distribution of the hydrogen columns of absorbed AGNs for 
different models and derived by [29].} \label{fig1} \vspace{0cm} 
\end{figure*}

\end{document}